%% file: paper.tex
\documentclass[]{trbunofficial}
\usepackage{graphicx}
\usepackage[table]{xcolor}

\usepackage[hidelinks]{hyperref}
\input{tex-helpers/macros}

\usepackage{caption}

\graphicspath{{./img/}}

\AuthorHeaders{Md Johirul Islam, Anuj Sharma and Hridesh Rajan}
\title{A Cyberinfrastructure for BigData Transportation Engineering}

\author{%
  \textbf{Md Johirul Islam}\\
  mislam@iastate.edu\\
  \hfill\break
  \textbf{Anuj Sharma}\\
   anujs@iastate.edu\\
  \hfill\break%
  \textbf{Hridesh Rajan}\\
   hridesh@iastate.edu
}

\TotalWords{4217}

\begin{document}
\maketitle

\section{Abstract}

\input{abstract}

\hfill\break %
\noindent\textit{Keywords}: Big Data, Domain-specific-language, Cyberinfrastructure
\newpage

\input{introduction}
\input{motivation}
\input{related}
\input{approach}
\input{evaluation}

\input{conclusion}

\section{Acknowledgements}
  This material is based upon work supported by the
   National Science Foundation under Grant CCF-15-18897 and CNS-15-13263.  
   Any opinions, findings, and conclusions or recommendations expressed 
   in this material are those of the authors and do not necessarily reflect the 
   views of the National Science Foundation.

\newpage

\bibliographystyle{trb}
\bibliography{refs}
\end{document}

%% file: tex-helpers/macros.tex
\usepackage{relsize}
\usepackage[T1]{fontenc}
\usepackage{times}
\usepackage{url}
\usepackage{amssymb}
\usepackage{amsmath}
\usepackage{rotating}
\usepackage{colortbl}
\usepackage{tabularx}
\usepackage[table]{xcolor}
\usepackage{xspace}

\usepackage{epsfig}
\usepackage{subfigure}
\usepackage{multirow}
\graphicspath{{figures/}}
\usepackage{graphicx}

\input{tex-helpers/obey}

\input{tex-helpers/grammar}

\usepackage{listings}

\usepackage{algorithm}
\usepackage{algorithmic}

\definecolor{lightgray}{gray}{0.97}
\definecolor{codeblue}{rgb}{0.13,0.13,1}
\definecolor{OliveGreen}{cmyk}{0.64,0,0.95,0.40}
\definecolor{DarkGreen}{cmyk}{0.40,0,0.50,0.15}
\definecolor{purple}{cmyk}{0.41,0.73,0,0}
\definecolor{LightRed}{cmyk}{0,0.682,0.728,0}

\makeatletter
\lst@Key{countblanklines}{true}[t]%
    {\lstKV@SetIf{#1}\lst@ifcountblanklines}

\lst@AddToHook{OnEmptyLine}{%
    \lst@ifnumberblanklines\else%
       \lst@ifcountblanklines\else%
         \advance\c@lstnumber-\@ne\relax%
       \fi%
    \fi}
\makeatother

\newcommand{\figref}[1]{Figure~\ref{#1}}
\newcommand{\secref}[1]{Section~\ref{#1}}

\newcommand{\etal}{~\textit{et al.}\@\xspace}

%% file: tex-helpers/obey.tex
{\catcode`\^^M=13 \gdef\myobeycr{\catcode`\^^M=13 \def^^M{\\}}%
}

{\catcode`\^^I=13 \gdef\obeytabs{\catcode`\^^I=13 \def^^I{\hbox{\hskip 4em}}}}

{\obeyspaces\gdef {\hbox{\hskip0.5em}}}


%% file: tex-helpers/grammar.tex
\newcommand{\goesto}{\mbox{$::=$}}

\newenvironment{grammar}{
  \def\:{\goesto{}}
  \def\|{$\vert$}
  \tt \myobeycr%
  \begin{tabbing}%
  \qquad \= $\vert$ \= \qquad \= \kill%
}%
{\unskip\end{tabbing}}

\newenvironment{grammar*}{
  \def\:{\goesto{}}
  \def\|{$\vert$}
  \tt%
  \begin{tabbing}%
  \qquad \= $\vert$ \= \qquad \= \kill%
}%
{\unskip\end{tabbing}}

\newenvironment{roman-grammar}{
  \def\:{\goesto{}}
  \def\|{$\vert$}
  \myobeycr%
  \begin{tabbing}%
  \qquad \= $\vert$ \= \qquad \= \hspace*{5cm} \= \qquad \= \kill%
}%
{\unskip\end{tabbing}}

\newenvironment{roman-grammar*}{
  \def\:{\goesto{}}
  \def\|{$\vert$}
  \begin{tabbing}%
  \qquad \= $\vert$ \= \qquad \= \hspace*{5cm} \= \qquad \= \kill%
}%
{\unskip\end{tabbing}}

{\unskip\end{displaymath}}

\newlength{\BEFOREGRAMCOMMENTLEN}
\newenvironment{commented-grammar}{%
\begin{grammar}%
\qquad \= $\vert$ \= \qquad \hspace*{\BEFOREGRAMCOMMENTLEN} \= \kill%
}{\end{grammar}}%

%% file: abstract.tex
Big Data-driven transportation engineering has the potential to improve 
utilization of road infrastructure, decrease traffic fatalities, improve fuel 
consumption, decrease construction worker injuries, among others. 
Despite these benefits, research on Big Data-driven transportation 
engineering is difficult today due to computational expertise required to 
get started. 
This work proposes BoaT, a transportation-specific programming language, and
its Big Data infrastructure that is aimed at decreasing this barrier to entry.
Our evaluation that uses over two dozen research questions from six 
categories show that research is easier to realize as a BoaT computer program, 
an order of magnitude faster when this program is run, 
and exhibits 12-14x decrease in storage requirements.

%% file: introduction.tex
\section{Introduction}
\label{sec:introduction}

The potential and challenges of leveraging Big Data in transportation has long been recognized \cite{jagadish2014big,lv2015traffic,seedah2015approach,zhang2011data,barai2003data,kitchin2014real,fan2014challenges,laney20013d,chen2014data,wang2017big,chakraborty2017outlier,liu2016data,huang2016leveraging,adu2017framework}.
For example, researchers have shown that Big Data-driven transportation engineering can 
help reduce congestions, fatalities and make building transportation applications easier \cite{zhang2011data,barai2003data,huang2016leveraging}.
The availability of open transportation data that is accessible e.g. on the web under a permissive license, has the potential 
to further accelerate the impact of Big Data-driven transportation engineering.

Despite this incredible potential, harnessing Big Data in transportation for research remains difficult. 
In order to utilize Big Data, expertise is needed along each of the five steps of a typical data pipeline 
namely data acquisition; information extraction and cleaning; data integration, aggregation, and representation;
modeling and analysis; and interpretation \cite{jagadish2014big}.
First three steps are further complicated by the heterogeneity of data from multiple sources \cite{seedah2015approach},
e.g. speed sensors, weather station, national highway authority.
A scientist must understand the peculiarities of the data sources to develop a data acquisition mechanism, clean data coming from multiple sources, and integrate data from multiple sources.
Modeling and analysis are complicated by the volume of the data.
For example, a dataset of speed measurements from a commercial provider for Iowa for 
a single day can be in multiple GBs, exceeding the limits of a single machine.
Analyses that aim to compute trends over multiple years can easily require storing, and computing over,  tens of TBs of just speed sensor data. 

%
A possible solution could be to use the Big data technologies like Hadoop, Apache Spark
running over a distributed cluster. 
Using a distributed cluster with an adequate number of nodes problems related to the storage and time of computation can be addressed. 
But these Big Data technologies are not so easy to use. 
Getting started requires technical expertise to set up the infrastructure, efficient design of data schema,
data acquisition strategy from multiple sources, high level of programming skills, adequate knowledge of
distributed computing models and a lot more efficiency in writing distributed computer programs which 
is significantly different than writing a sequential computer program in Matlab, C or Java. 
The analysis of Big Data in transportation is almost an elite job due to these barriers. 
The research groups interested in Big Data-driven transportation engineering have to hire technically skilled people or train their own staff members to use these highly sophisticated technologies. 
Both approaches incur additional costs. 

This work describes a transportation-specific Big Data programming language and its infrastructure that is aimed at solving these problems. 
We call this language BoaT (Boa\cite{Dyer2015} for Transportation). 
The BoaT infrastructure provides build-in transportation data schemas and converters from existing data sources. 
A notable advantage of BoaT's data schema is a significant reduction in storage requirements. 
A transportation researcher or engineer can express their queries as simple sequential looking BoaT programs
that is another advantage of the approach.
The BoaT infrastructure automatically converts a BoaT program to a distributed executable code without sacrificing correctness in the conversion process. 
This also often results in an order of magnitude improvement in performance that is the third advantage of our approach. 
The BoaT infrastructure provides build-in transportation data schemas and converters from existing data sources. 
The four notable advantages of BoaT are : 
a.) significant reduction in storage requirement by using specially designed data schema, 
b.) A transportation researcher or engineer can express their queries as simple sequential looking BoaT programs, 
c.) auto conversion of sequential programs to a parallelly executable programs without sacrificing correctness in the conversion process, 
d.) The number of lines of code significantly reduces thus reducing the 
debugging time for the program. 
Owing to these advantages, even users that are not experts in distributed 
computing can write these BoaT programs that lower the aforementioned barrier to entry.

The remainder of this article describes the BoaT approach and explores its advantages. 
First, in the next section, we motivate the approach via a small example. 
Next, we compare and contrast this work with related ideas.
Then, we describe the salient technical aspects of the technique. 
Next, we evaluate the usability, and scalability of the technique, show some example use cases that 
we have realized, and highlight benefits of our storage strategy.
Finally, we conclude.

%% file: motivation.tex
\section{Motivation}
\label{sec:motivation}

Transportation agencies collect a lot of data to make critical data driven decision for Intelligent Transportation System (ITS). 
There has been a lot of initiative to make data available for researchers to spur innovation \cite{opendata}. 
But the analysis of this ultra large-scale data is a difficult task given the technology needed to analyze the data is still a luxury \cite{biuk2016big}. 
These data come from multiple sources with a lot of varieties, 
velocities, and volumes. 
Given the availability of a variety of sources of data technically skilled people also often face challenges due to the kinds of input, data access patterns, type of parallelism, etc. \trbcite{kambatla2014trends}. 
The need to write complex programs can be a barrier for domain 
researchers to take the advantage of this large-scale data.
To illustrate the challenges, consider a sample question ``Which 
counties have highest and lowest average temperature in a day?'' A query like this is simple when the data is already provided by county but in case you have data for every 5 minute for every square mile of Iowa for last 10 years. The query becomes hard to solve in Matlab or even R and could potentially run for a long time in Java
Answering this question in Java would require knowledge of (at a minimum):
reading the weather data from the data provider service,
finding the locations and county information of different grids from some other APIs, additional filtering code, controller logic, etc. 
Writing such a program in Java, for example, would take upwards of 
100 lines of code and require knowledge of at least 2 complex libraries and 2 complex data structures. 
A heavily elided example of such a program is shown in \figref{fig:javaboa}, 
left column. 

\input{examples/nmotiv}

This program assumes that the user has manually downloaded
the required weather data, preprocessed the data and written to a CSV file. 
It then processes the data and collects weather information in different grids at different times of the day. 
Next, the county information of each grid is found from another API.
Finally, the data is stored in some data structures for further computation. 
The presented program is sequential and will not scale as the
data size grows. One could write a parallel computation program which would be even more complex.

We propose a domain-specific programming language called BoaT to solve these problems. 
We intend to lower the barrier to entry and enable the analysis of ultra-large-scale transportation data for answering more critical data-intensive research challenges. 
The main features of Boa for transportation data analysis originated from \cite{dean2008mapreduce,Dyer2015,pike2005interpreting,urso2012sizzle}.
To this, we add builtin transportation specific data types and functions for 
analysis of large-scale transportation data, schema, and infrastructure 
to preprocess data automatically and store efficiently.
The main components come as an integrated framework that provides 
a domain specific language for transportation data analysis, a data processing unit, and a storage strategy.

%% file: examples/nmotiv.tex

\newcommand*{\vcenteredhbox}[1]{\begingroup
	\setbox0=\hbox{#1}\parbox{\wd0}{\box0}\endgroup}

\begin{sidewaysfigure}
	\begin{center}
		\begin{minipage}[t]{0.44\textwidth}
			\begin{center}
				\textbf{Java}
			\end{center}
			\vspace{-1em}
			\input{examples/motivjava}
		\end{minipage}
		\begin{minipage}[t]{0.55\textwidth}
			\begin{center}
				\textbf{BoaT}
			\end{center}
			\vspace{-1em}
			\input{examples/motivboa}

			\begin{center}
				\textbf{{\em Performance}}\\
				\vspace{0.5em}
				\includegraphics[width=\linewidth]{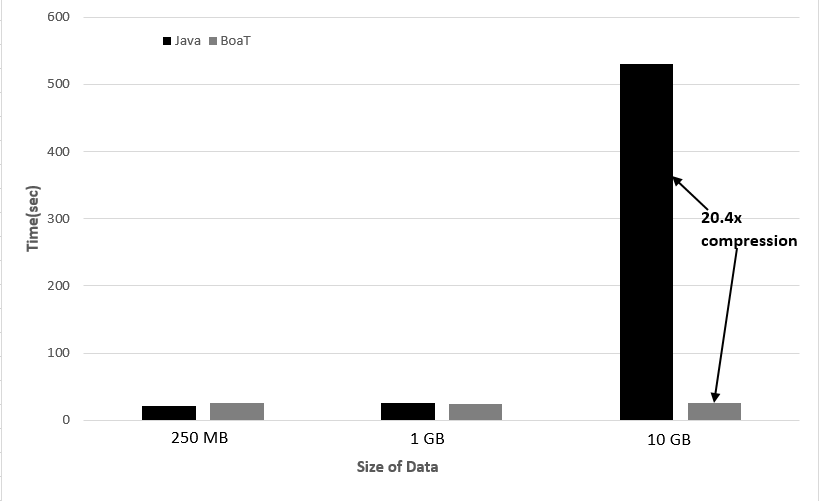}
			\end{center}
		\end{minipage}
		\caption{Programs for answering "Which counties have highest and lowest average temperature in a day?" and the performance with the size of data}
		\label{fig:javaboa}
	\end{center}
\end{sidewaysfigure}

%% file: examples/motivjava.tex
\begin{lstlisting}[language=Java,firstnumber=1]
	... // imports
\end{lstlisting}
\vspace{-\baselineskip}
\newcount\skipLinesCnt\skipLinesCnt=\numexpr8\relax 
\begin{lstlisting}[language=Java,firstnumber=\skipLinesCnt]
	public class CountiesMaxAvgTmpc {
	public static void main(String[] args){
	... // File operation
	PriorityQueue<CountyTemp> maxheap , minheap;
	... // data processing
\end{lstlisting}
\vspace{-\baselineskip}
\skipLinesCnt=\numexpr25+40\relax 
\begin{lstlisting}[language=Java,firstnumber=\skipLinesCnt]
	private String buildHeaps(/*input heaps*/) {
	... // Populate the heaps with Objects
	}
\end{lstlisting}
\vspace{-\baselineskip}
\skipLinesCnt=\numexpr12+60\relax 
\begin{lstlisting}[language=Java,firstnumber=\skipLinesCnt]	
	... // Iterating over the Map to find average temperature of each county
	for (Map.Entry<String, List<Double>> entry : map.entrySet()) {
		String county = entry.getKey();
		List<Double> countyTemps = entry.getValue();
		for(double temp : countyTemps){
			...// Iterating over the county temp for avg
		}


\end{lstlisting}
\vspace{-\baselineskip}
\skipLinesCnt=\numexpr10+79\relax 
\begin{lstlisting}[language=Java,firstnumber=\skipLinesCnt]
	// Getting County name from another data file given the grid id
	public static String getCounty(int gridid) {
	...	// Code to find county from grid
	}
	// Internal class to hold county data
	private static class CountyTemp {
	... // County variables
	public CountyTemp(String countyName, double temperature) {
		... // Constructor
	}	
	... // codes to manage county data
	}
	public static String getCounty(int gridid)  {
	... // Getting County from a  grid id from API
		String county = getCountNameFromAPI(gridid);
\end{lstlisting}

%% file: examples/motivboa.tex
\begin{lstlisting}
	p: County = input;@\label{line:input}@
	max: output maximum(1) of string weight float;@\label{line:output}@
	min: output minimum(1) of string weight float; @\label{line:output1}@
	count := 0;
	sum := 0.0;
	visit(p, visitor{ @\label{line:visit}@
		before n: Grid -> {
		weatherRoot := getweather(n,"5-11-2017");
		foreach(s : int; def(weatherRoot.weather[s])) {
		sum = sum + weatherRoot.weather[s].tmpc;
		count++;
		}
		}
		});
	max << p.countyName weight sum/count; @\label{line:max}@
	min <<  p.countyName weight sum/count; @\label{line:min}@
\end{lstlisting}

%% file: related.tex
\section{Related Work}
\label{sec:related}

Due to the rapid growth of data-driven Intelligent transportation system 
(ITS) \trbcite{el2011data} \trbcite{zheng2015methodologies} applications 
and smart cities the necessity of harnessing the power of ultra-large-scale 
data is becoming more important today than any time before. 
Though a lot of works are done on data-driven smart city design and 
Big data analysis, trying to tackle the challenges of transportation big data from the domain-specific language perspective is few. 
In other domains, a lot of advantages are being taken from Big data using 
Domain Specific programming languages.
For example, \trbcite{Dyer2015} used the early version of Boa to analyze ultra-large-scale software repository data (data from repositories like GitHub). 
However, Dyer \etal's work is limited to software repositories whereas 
BoaT built on top of Boa is provides the support of transportation data analysis at ultra-large scale, transportation domain types and an infrastructure of efficient data storage from a variety of transportation data sources.

There has been some efforts to support domain types and computation in 
transportation in an integrated modeling tool called  
UrbanSim \trbcite{borning2008domain}, \trbcite{borning2008urbansim}, 
\trbcite{waddell2003microsimulation}. 
UrbanSim is an integrated modeling environment that provides a modeling language which provides access to urban data for finding models to coordinate transportation and land usage \trbcite{waddell2003microsimulation}.
While UrbanSim focuses on simulation, BoaT is for analyzing gathered data. 
Furthermore, supporting analysis of large-scale data has not been the focus of 
UrbanSim, whereas BoaT focuses on providing scalable support for data analysis.
\trbcite{simmhan2013cloud} provides a cloud-based software platform for data analytics in Smart Grids, whereas BoaT is focussed on transportation data.
%
\trbcite{du2016active}'s City Traffic Data-as-a-Service (CTDaaS) uses service-oriented 
architecture to provide access to data, but does not focus on the scalable analysis of Big Data.

In general, the current approaches using big data analytics are either using costly cloud computation or have custom build design for solving specific problems using open source solution with on-premise servers. 
Works such as \trbcite{yang2015big} and \trbcite{wang2016traffic} highlight
the challenges of doing Big Data-driven transportation engineering today. 
For example, \trbcite{yang2015big} use HDFS, MLlib, cluster computing to solve their problems, essentially like our motivating example. 
Each of these technologies creates its own barrier to entry.
There is a need for a framework that would overcome the barrier to use big data analytics, 
provide a domain specific language, reduce the efforts of data preprocessing and will be 
available at a mass scale. 

%% file: approach.tex
\section{BoaT: Design and Implementation}
\label{sec:approach}

\begin{figure}[h]
	\includegraphics[width=\textwidth]{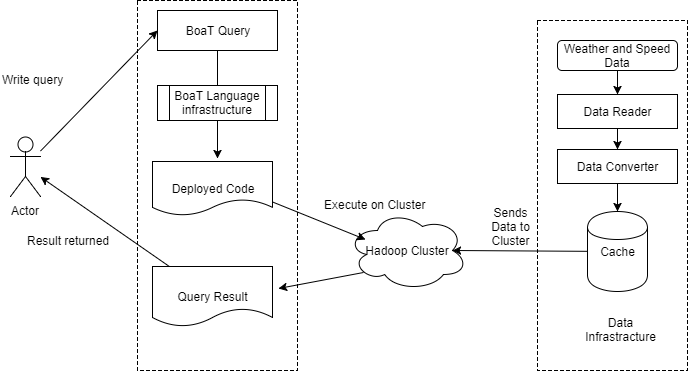}
	\caption{An Overview of BoaT: shows workflow of a BoaT user 
		and BoaT infrastructure}   
	\label{fig:boa}
\end{figure}

To address the challenges of easy and efficient analysis of big transportation data we propose a transportation-specific programming language and data infrastructure. 
The language provides simple syntax, domain-specific types and massive abstractions. 
An overview of the infrastructure is shown in \figref{fig:boa}.

The user writes the BoaT program and submits it to the BoaT infrastructure, 
The BoaT program is taken by the infrastructure and converted by a specialized compiler that we have written to produce an executable that can be deployed in a distributed Hadoop cluster. 
This executable is run automatically on curated data to produce output for the user.

To illustrate, consider the question in \secref{sec:motivation} 
``Which counties have the highest and the lowest average temperatures in a day?''.
A BoaT program to answer this question is shown in \figref{fig:javaboa},
right column. Line 1 of the program says that it takes a County as input.
So, if there are n counties in the dataset, the statements on lines 4-16 of this program would be automatically run in parallel by the BoaT infrastructure (once for each county). 
Line 2 and 3 of this program declares output variables. 
These {\em write only} output variables are shared between all parallel tasks 
created by the BoaT infrastructure and the infrastructure manages the details
of effectively interleaving and maximizing performance.
Line 2 says that this output variable will collect values written to it and compute
the maximum of those values. This is called {\em aggregation} in BoaT and several
other kinds of aggregation algorithms are supported as shown in \figref{tbl:agg}.
Line 15 shows an example of writing to that output variable. 
Lines 4-16 are run sequentially for each county. 
They look into each grid of the county (lines 6,7,14) to find temperature data of the grid while maintaining a running sum and frequency to compute average on 
lines 15-16. 
While the details of this program are important also, astute readers would have surely observed that writing this program needed no knowledge of how the data is accessed, what is the schema of the data, how to parallelize the program. 
No parallelization and synchronization code is needed. 
The BoaT program produces result running in a Hadoop cluster. 
So the program scales well saving hours of execution time. 

As the program runs on a cluster it outperforms the Java program (sequential) 
as the input data size grows. 
A comparison is shown in \figref{fig:javaboa} on the lower right corner. 
The BoaT program provides output almost 20.4 times faster only on one-day
weather data of Iowa (10GB). 
To achieve these goals we have solved following problems.
\begin{itemize}
	\item Providing transportation domain types and functions;
	\item designing the schema for efficient storage strategy and parallelization; and
	\item providing an effective solution to data fusion.
\end{itemize}


\subsection{Language Design}
\label{subsec:language}

The language BoaT is the extended version of work done by \trbcite{Dyer2015}. 
They provide the syntax and tools to analyze the mining software repository data. 
We extended their work to provide domain types, functions and computational infrastructure for Big Data-driven transportation engineering. 
We create the schema using Google protocol buffer. 
Google protocol buffer is an efficient \trbcite{Dyer2015} data representation format that provides faster memory efficient computation in BoaT.

\subsection{Domain Types}
\label{subsec:domain-types}

\input{tables/domaintypes}
\input{tables/aggregator}

The transportation-specific types in BoaT are shown in \figref{tbl:types}.
As we and others use this infrastructure these types will surely evolve, and the BoaT infrastructure is designed to support such evolution.
County is the top level type. 
This type has attributes that relate to the code of the county, name of the county and a list of grids in the county. 
A grid is related to a location in a county. For the convenience of computation, the whole Iowa is split into 213840 Grids by Iowa DOT. 
So we also used Grid as the domain. 
The Grid has attributes Id, reference to the WeatherRoot which refers to the weather records in that Grid, reference to SpeedRoot which refers to the Speed records in that Grid. 
Weather root contains a list of weather records. 
SpeedRoot contains a list of Speed records. 
So we can easily go to the speed or weather data of a particular location in a particular Grid under a particular County without searching through all the data in the cluster. 
SpeedRecord contains the attributes detectorCode, type of detector, average, 
reference, roadname and time.

The data design has led to two innovations. 
First to balance query speed, flexibility, and storage capacity. 
Second to allow future extension via data fusion. 

While designing the schema we came to a successful data reduction strategy after multiple trials. 
Initially, we were using all the data at the top level. 
That means when we access a row we accessed all the relevant data for that row like weather, speed.  
Following this strategy, the storage size increased than the raw data. 
Then we split the data keeping county data at the top level and the relevant weather, speed records at the second level in the same list. 
We were not getting enough mappers to make a lot of parallelization in the program as the splitting was not possible. 
And at the same time storage size was almost near the raw data size. 
Then we made multiple levels of hierarchy in our type system. 
The top level is the county. 
The county contains a list of grids (spatial locations), each grid contains two optional fields to point to speed data and weather data. 
This strategy of data representation gives us benefit in storage as well as in faster computation as only relevant data is accessed.  
We can store incremental data without regenerating the whole dataset from the beginning. 
Without this hierarchical schema strategy, all the data need to be merged together creating a merged schema hampering the sustainability, 
scalability and storage benefit of the system. 
And the addition of new data would be impossible.

Fusion of multiple data sources in existing big data frameworks is difficult due to size, the necessity of join and parallel queries in the data sources. 
In BoaT, we addressed this problem in data infrastructure. 
Any new dataset can be added to the infrastructure easily. 
For example, we started with speed dataset initially and we were able to answer questions on speed data. 
The access link to speed data is optional. 
That means we don't load the data unless it is necessary. 
Then we added another optional link to weather dataset. 
We came up with a successful fusion of data and were able to answer queries that cover both speed and weather dataset without losing
any performance. 
The queries of category E in \figref{tbl:app-result} are examples of using the fusion of weather and speed dataset. 
And the performance is not affected by this. 
This makes our infrastructure sustainable to any new datasets of interest to be added to the infrastructure. 
To do that we have to just add an optional link to that new dataset after providing the schema for new dataset. 
The infrastructure will take care of all other complexities related to 
data generation, and type generation.





%% file: tables/domaintypes.tex

\begin{figure*}[ht]
\centering
\rowcolors{2}{gray!25}{white}
\begin{footnotesize}
\begin{tabular}{|l||r|r|r||r|r|r|}
\cline{2-7}
\hline
\hline
\multicolumn{1}{||c||}{\textbf{Type}} & \multicolumn{3}{c||}{\textbf{Attributes}} & \multicolumn{3}{c|}{\textbf{Details}} \\
\hline
\hline
\multicolumn{1}{||c||}{} & \multicolumn{3}{c||}{countyCode} & \multicolumn{3}{c|}{Code of the county}
\\
\multicolumn{1}{||c||}{\textbf{County}} & \multicolumn{3}{c||}{countyName} & \multicolumn{3}{c|}{Name of the county}
\\
\multicolumn{1}{||c||}{} & \multicolumn{3}{c||}{Grids} & \multicolumn{3}{c|}{List of Grid in the county.}
\\
\hline
\hline
\multicolumn{1}{||c||}{} & \multicolumn{3}{c||}{ID} & \multicolumn{3}{c|}{ID of a grid}
\\
\multicolumn{1}{||c||}{\textbf{Grid}} & \multicolumn{3}{c||}{Location} & \multicolumn{3}{c|}{Spatial location of the grid}
\\
\multicolumn{1}{||c||}{} & \multicolumn{3}{c||}{WeatherRoot} & \multicolumn{3}{c|}{Link to the Weather data for the grid}
\\

\multicolumn{1}{||c||}{} & \multicolumn{3}{c||}{SpeedRoot} & \multicolumn{3}{c|}{Link to the speed data for that grid}
\\
\hline
\hline
\multicolumn{1}{||c||}{\textbf{SpeedRoot}} & \multicolumn{3}{c||}{speedRecords} & \multicolumn{3}{c|}{List of SpeedRecord}
\\
\hline
\multicolumn{1}{||c||}{\textbf{WeatherRoot}} & \multicolumn{3}{c||}{weatherRecords} & \multicolumn{3}{c|}{List of WeatherRecord}
\\
\hline
\hline
\multicolumn{1}{||c||}{} & \multicolumn{3}{c||}{detectorcode} & \multicolumn{3}{c|}{The code of the detector giving the current record}
\\
\multicolumn{1}{||c||}{} & \multicolumn{3}{c||}{type} & \multicolumn{3}{c|}{Type of the vehicle}
\\
\multicolumn{1}{||c||}{\textbf{SpeedRecord}} & \multicolumn{3}{c||}{speed} & \multicolumn{3}{c|}{Speed of the vehicle}
\\
\multicolumn{1}{||c||}{} & \multicolumn{3}{c||}{reference} & \multicolumn{3}{c|}{Reference speed}
\\
\multicolumn{1}{||c||}{} & \multicolumn{3}{c||}{time} & \multicolumn{3}{c|}{Time of the record}
\\
\multicolumn{1}{||c||}{} & \multicolumn{3}{c||}{roadname} & \multicolumn{3}{c|}{Name of the road of the record}
\\
\hline
\hline
\multicolumn{1}{||c||}{} & \multicolumn{3}{c||}{tmpc} & \multicolumn{3}{c|}{2 m above the ground level temperature}
\\
\multicolumn{1}{||c||}{} & \multicolumn{3}{c||}{wawa} & \multicolumn{3}{c|}{Watches, warnings, and advisories issued by the National Weather Service}
\\
\multicolumn{1}{||c||}{} & \multicolumn{3}{c||}{ptype} & \multicolumn{3}{c|}{Type of Precipitation}
\\
\multicolumn{1}{||c||}{} & \multicolumn{3}{c||}{dwpc} & \multicolumn{3}{c|}{Dew point temperature}
\\
\multicolumn{1}{||c||}{} & \multicolumn{3}{c||}{smps} & \multicolumn{3}{c|}{Wind speed}
\\
\multicolumn{1}{||c||}{\textbf{WeatherRecord}} & \multicolumn{3}{c||}{drct} & \multicolumn{3}{c|}{Wind direction}
\\
\multicolumn{1}{||c||}{} & \multicolumn{3}{c||}{vsby} & \multicolumn{3}{c|}{Horizontal visibility from sensors in Km}
\\
\multicolumn{1}{||c||}{} & \multicolumn{3}{c||}{roadtmpc} & \multicolumn{3}{c|}{Pavement surface temperature}
\\
\multicolumn{1}{||c||}{} & \multicolumn{3}{c||}{srad} & \multicolumn{3}{c|}{Solar radiation}
\\
\multicolumn{1}{||c||}{} & \multicolumn{3}{c||}{snwd} & \multicolumn{3}{c|}{Snow fall depth}
\\
\multicolumn{1}{||c||}{} & \multicolumn{3}{c||}{pcpn} & \multicolumn{3}{c|}{Precipitation accumulation}
\\
\multicolumn{1}{||c||}{} & \multicolumn{3}{c||}{time} & \multicolumn{3}{c|}{Time of the reading}
\\

\hline
\end{tabular}  
\end{footnotesize}
\caption{Domain types for transportation data in BoaT}
\label{tbl:types}
\end{figure*}

%% file: tables/aggregator.tex

\begin{figure*}[ht]
\centering
\rowcolors{2}{gray!25}{white}
\begin{footnotesize}
\begin{tabular}{|l||r|r|r||r|r|r|}

\cline{2-7}
\hline

\multicolumn{3}{||c|}{\textbf{Aggregator}} & \multicolumn{4}{c|}{\textbf{Description}} 
 \\
\hline
\hline
 \multicolumn{3}{|c|}{MeanAggreagtor} & \multicolumn{4}{c|}{Calculates the average}
\\
 \multicolumn{3}{|c|}{MaxAggreagtor} & \multicolumn{4}{c|}{Finds the maximum value}
\\
 \multicolumn{3}{|c|}{QuantileAggregator} & \multicolumn{4}{c|}{Calculates the quantile. An argument is passed to tell the quantile of interest}
\\
 \multicolumn{3}{|c|}{MinAggregator} & \multicolumn{4}{c|}{Finds the minimum value}
\\
 \multicolumn{3}{|c|}{TopAggregator} & \multicolumn{4}{c|}{Takes an integer argument and returns that number of top elements}
\\
 \multicolumn{3}{|c|}{StDevAggregator} & \multicolumn{4}{c|}{Calculates the standard deviation}
\\
\hline
\end{tabular}  
\end{footnotesize}
\caption{Aggregators in BoaT}
\label{tbl:agg}
\end{figure*}

%% file: evaluation.tex
\section{Evaluation and Results}
\label{sec:evaluation}

This section evaluates {\em applicability}, {\em scalability}, 
and {\em storage efficiency} of BoaT and its infrastructure. 
By applicability we mean whether a variety of transportation analytics
use cases can be programmed using BoaT. 
By scalability we mean whether the resulting BoaT programs scale when more resources are provided. 
By storage efficiency we mean whether storage requirements for data
are comparable to the raw data, or whether BoaT requires less storage, and if so how much. 

\input{tables/querytable}

\subsection{Applicability}
\label{subsec:applicability}

To support our claim of applicability we use BoaT to answer queries on weather and speed data to provide answers to multiple queries from different categories and classes. 
A small BoaT program can answer queries that would need a lot of efforts with other general purpose languages, distributed system and data processing. 
We provide a range of queries in six different categories and four different 
classes in table shown in \figref{tbl:app-result}. 

As an example scenario, we consider that a researcher wants to know the maximum and minimum temperature in different counties of a 
state in the USA in a date in May 2017. 
To achieve the result in the above scenario we have to write a small program in \figref{fig:example1}. 
All the complex technical details of Big Data analytics are abstracted from the user.  
Lets go through the program to understand what this small program is doing. 
In Line 1 we are taking the data as input. 
In our BoaT infrastructure, we currently use county as the top level entry point. 
In Line 2 and Line 3 we are declaring two output variables. 
The declaration tells clearly that one variable is going to store the maximum of some floating point numbers having a String i.e. the county name as key and the other variable is going to store the minimum of some floating point numbers. 
The floating point numbers here are temperature found from the data. 
In the next line there is a loop to iterate over all the grids of the county and for each county, we assign the temperature at that grid as weight. 
The program keeps track of the temperature values for each county and at 
the end returns maximum and minimum temperature at different 
counties in a day.

\input{examples/example1}

\begin{figure}[h]
	\includegraphics[width=\textwidth]{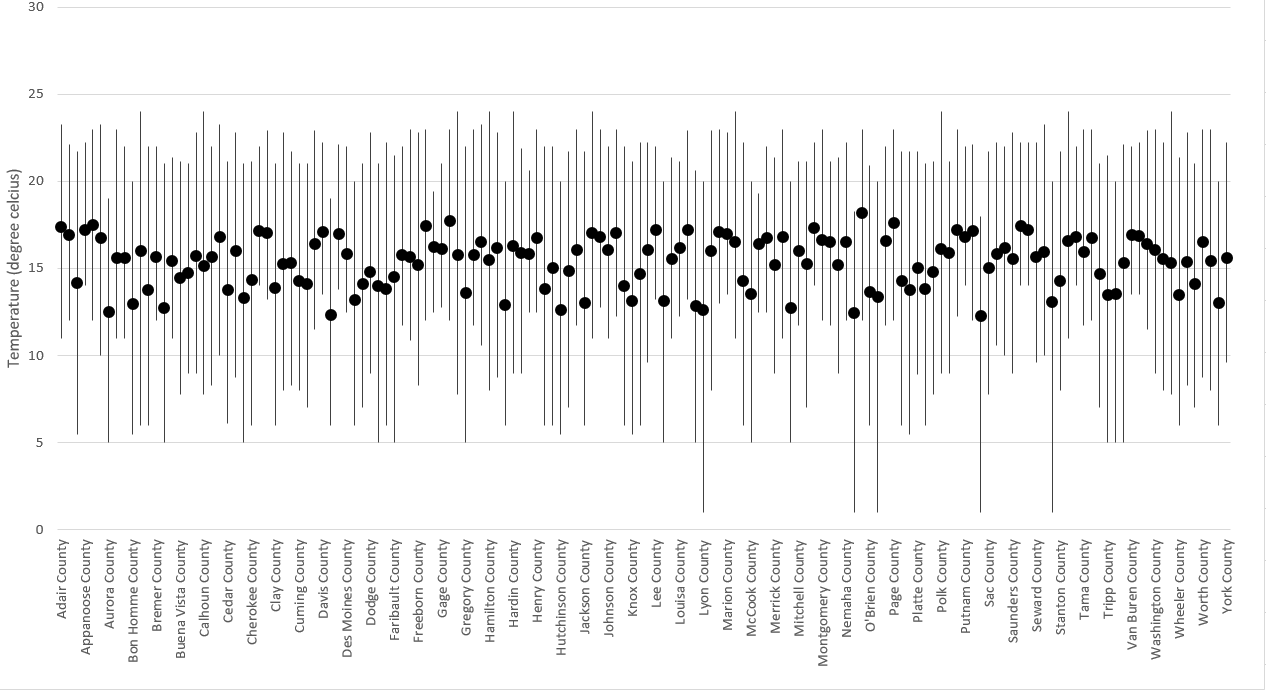}
	\caption{Error Bar graph of temperature showing minimum, maximum and average temperature of different  counties in a day. The result is produced from the code in \figref{fig:example1} and average is found from task A.1}   
	\label{fig:outputexample1}
\end{figure}

The output of the program is shown in \figref{fig:outputexample1}. \figref{fig:outputexample1} also contains average temperature which is 
computed from the task A.1. 

To go through another example lets take the task D.1. 
Here we calculate the mean and standard deviation of speed at 
different locations. 
The program is given in \figref{fig:example2}

\input{examples/example2}

The program like the program in \figref{fig:example1} first declares 
the output types. 
The output variable for mean uses the MeanAggregator in BoaT and 
the output variable for standard deviation uses the StDevAggregator. 
The program iterates through each county one by one and all the grids in that county. 
While visiting a grid of the county the program gets the speed data at that grid by using a domain specific function \texttt{getspeed()}. 
The function \texttt{getspeed()} has multiple versions and the version that we are using in this program takes the grid and a date as input and returns the speed data of that grid on that day. 
Then for each record of the speed data, we aggregate the values in the output. 
These visits run in different mapper nodes and the aggregation is 
done in different reducer nodes. 
Finally, the result is returned to the user.

We use two metrics to evaluate BoaT's applicability.
\begin{itemize}
	\item LOC: Line of Code. The total lines needed to write the program
	\item RTime: Runtime of the program
\end{itemize}

We show the comparison of these metrics for different programs in \figref{tbl:app-result}. 
The Java column shows the metric for Java program and the BoaT 
columns shows the values of the metrics for equivalent BoaT programs. 
The diff column shows how many times the BoaT program is efficient 
compared to Java in terms of Line of code. 
These Java programs are only for sequential operation. 
The Hadoop version of these programs can also be written, but that 
would require additional expertise and significantly larger lines of code. 

\begin{figure}%
	\centering
	\subfigure[Box plot of Lines of Code of Java and Boa]{%
		\label{fig:first}%
		\includegraphics[height=2in]{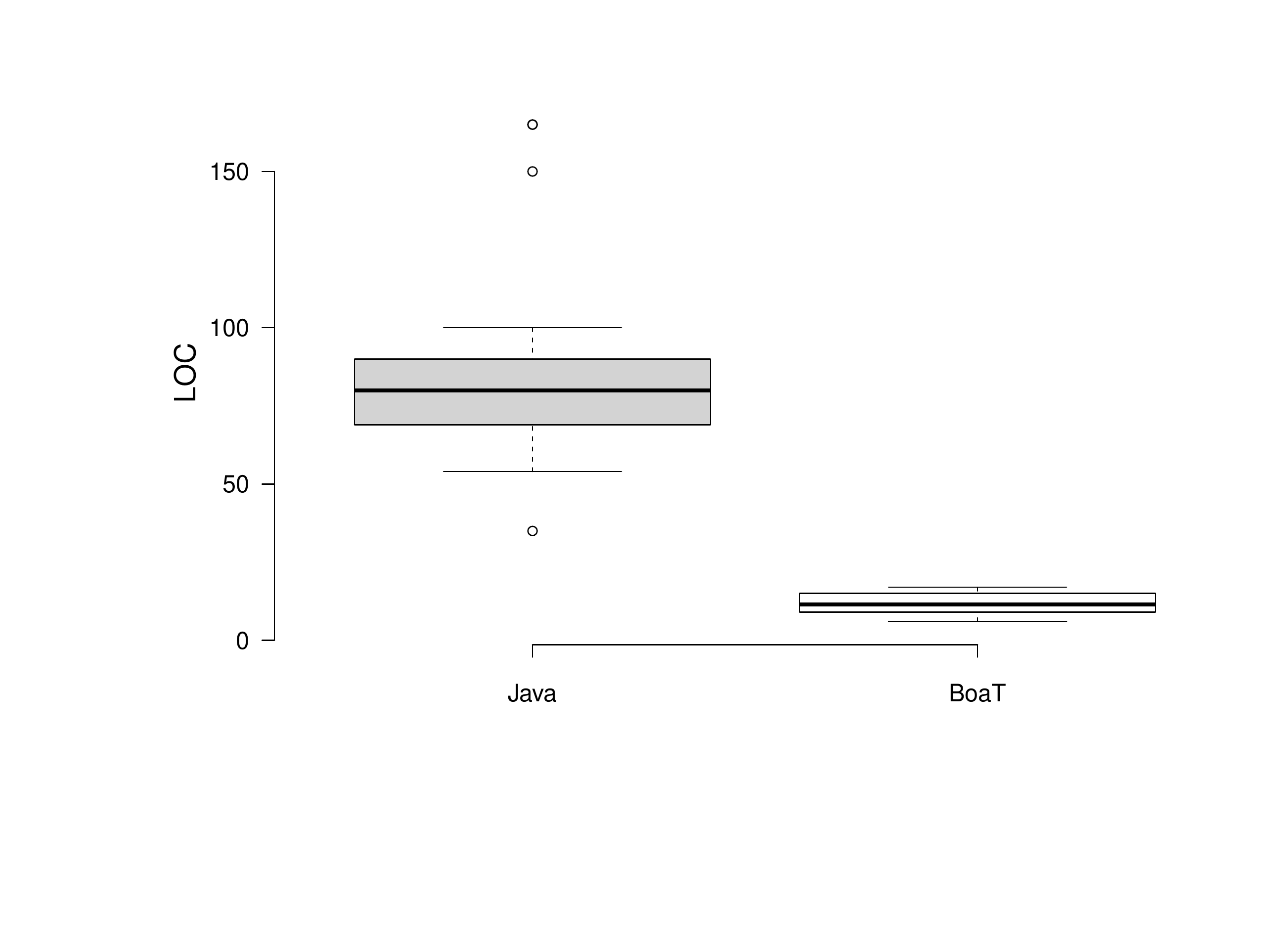}}%
	\qquad
	\subfigure[Box plot of RTime of Java and Boa]{%
		\label{fig:second}%
		\includegraphics[height=2in]{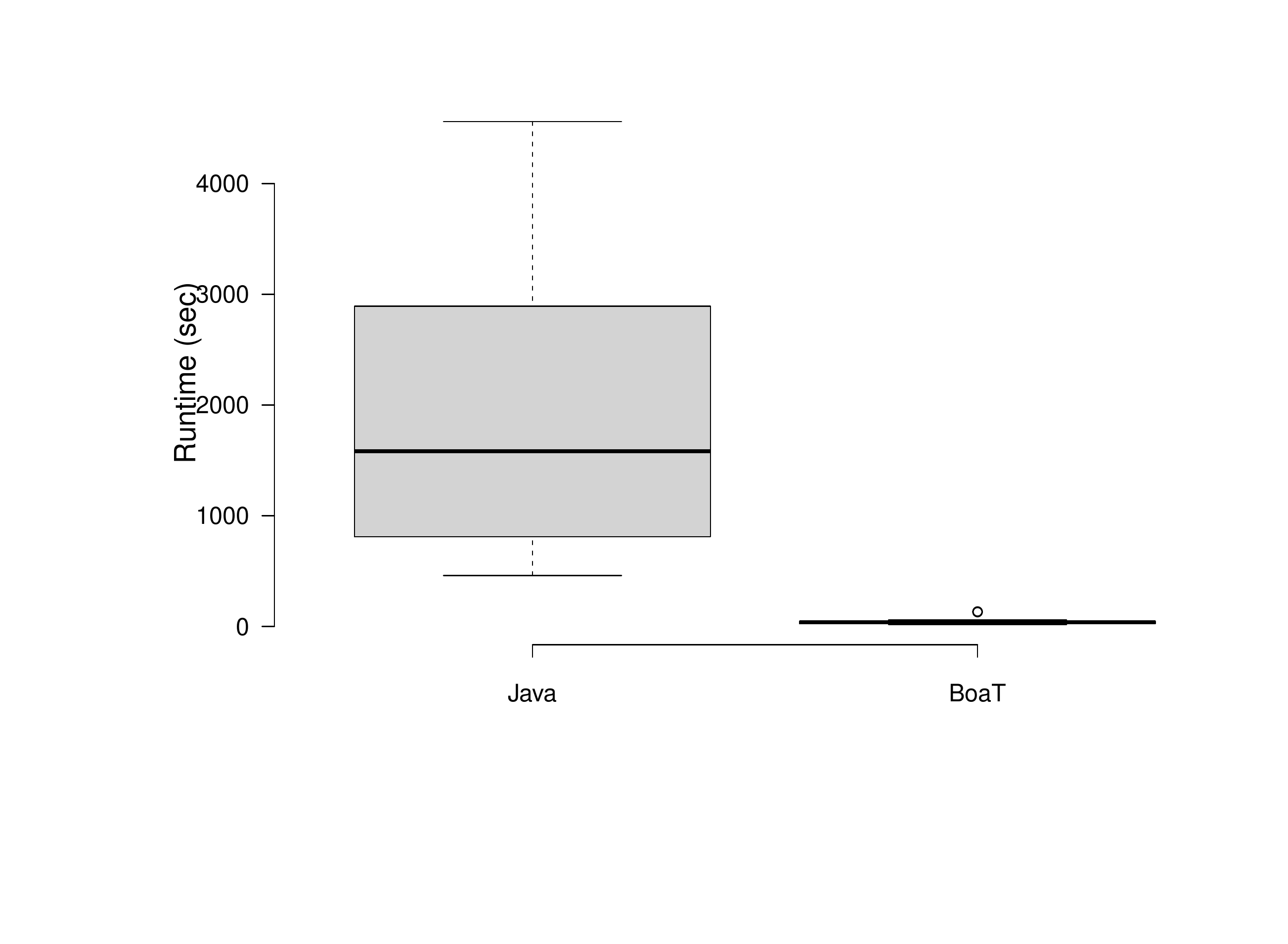}}%
	\caption{Lines of Code and Run time comparison between Java and BoaT Codes}
	\label{fig:rtloc}
\end{figure}

\subsection{Scalability}
\label{subsec:scalability}

Now we evaluate the scalability of BoaT programs. 
The compiled BoaT program runs in a Hadoop cluster. 
So BoaT provides all the advantages of parallel and distributed computation 
to the users that a Hadoop user would get. 

\begin{figure}[ht]
	\includegraphics[width=\textwidth]{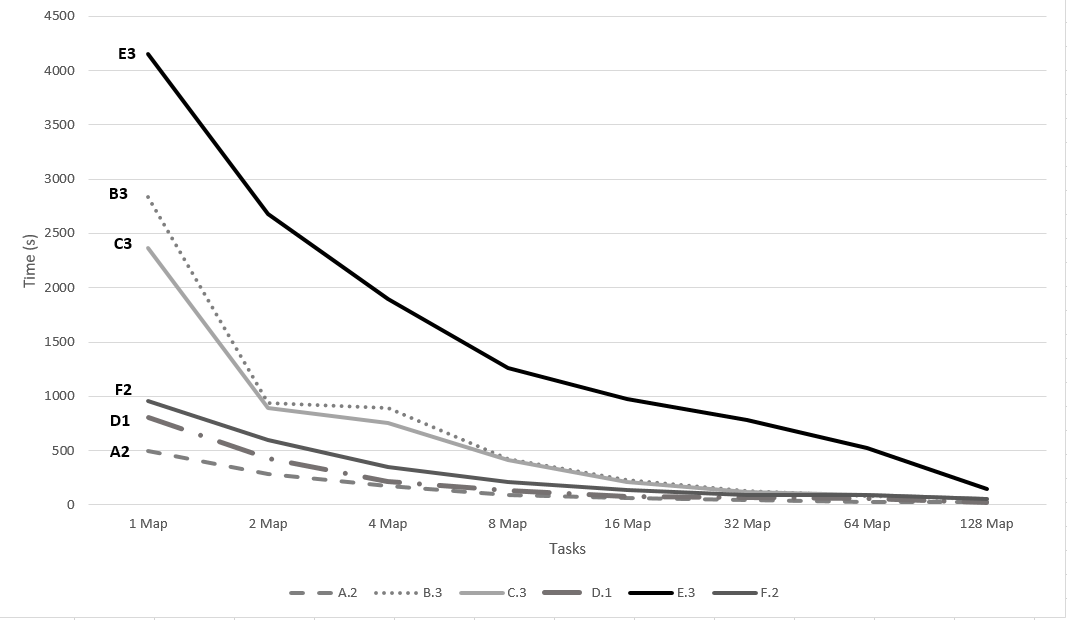}
	\caption{Scalability of BoaT programs. The trends show that BoaT program are 
		able effectively leverage the underlying infrastructure.}   
	\label{fig:scalability}
\end{figure}

To evaluate scalability we set up a Hadoop cluster with 23 nodes and with 
a capability of running 220 map tasks. 
We select one BoaT program from each category in \figref{tbl:app-result}. 
Then we run the programs gradually increasing number of map tasks. 
The result of running the programs is shown in \figref{fig:scalability}. 
The vertical axis represents the time in seconds. 
We see as the number of maps increases the run time of the program decreases. 

\begin{figure}%
	\centering
	\subfigure[Markers show the location of different speeding incidents. Once a marker is clicked the chart on the right side shows the number high-speed incidents categorized by speeds. This visualization is created from the result of the Task F.1]{%
		\label{fig:highspeedlocations}
		\includegraphics[width=.8\textwidth]{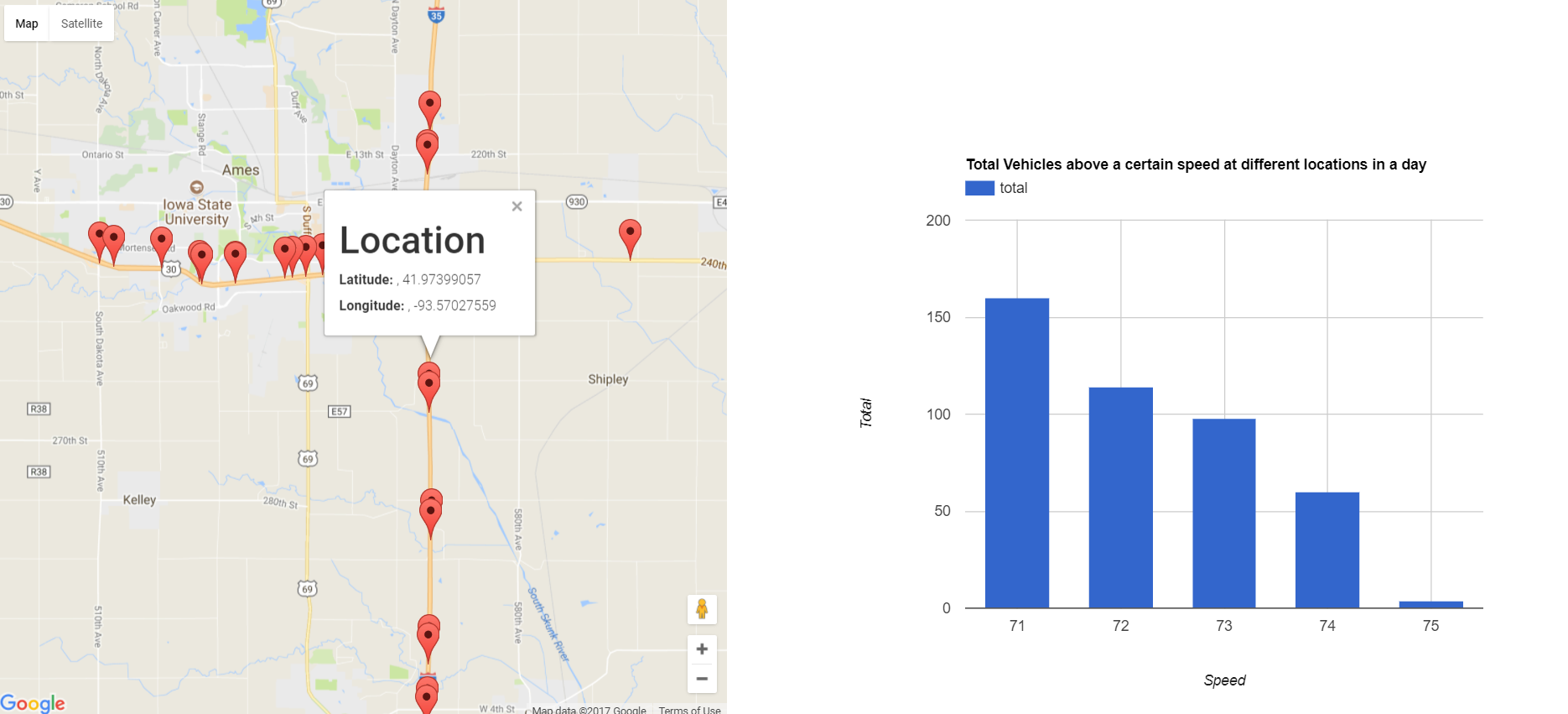}}%
	\newline
	\subfigure[Chart shows the counties with higher average speed on a day. This visualization is created from result of the Task D.2]{%
		\label{fig:topten}%
		\includegraphics[width = .8\textwidth]{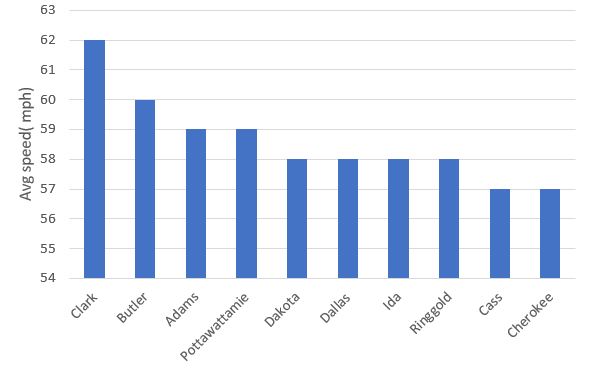}}%
	\caption{Visualization of tasks F.1 and D.2}
	\label{fig:speedviz}
\end{figure}
\begin{figure}[ht]
	\includegraphics[width=\textwidth]{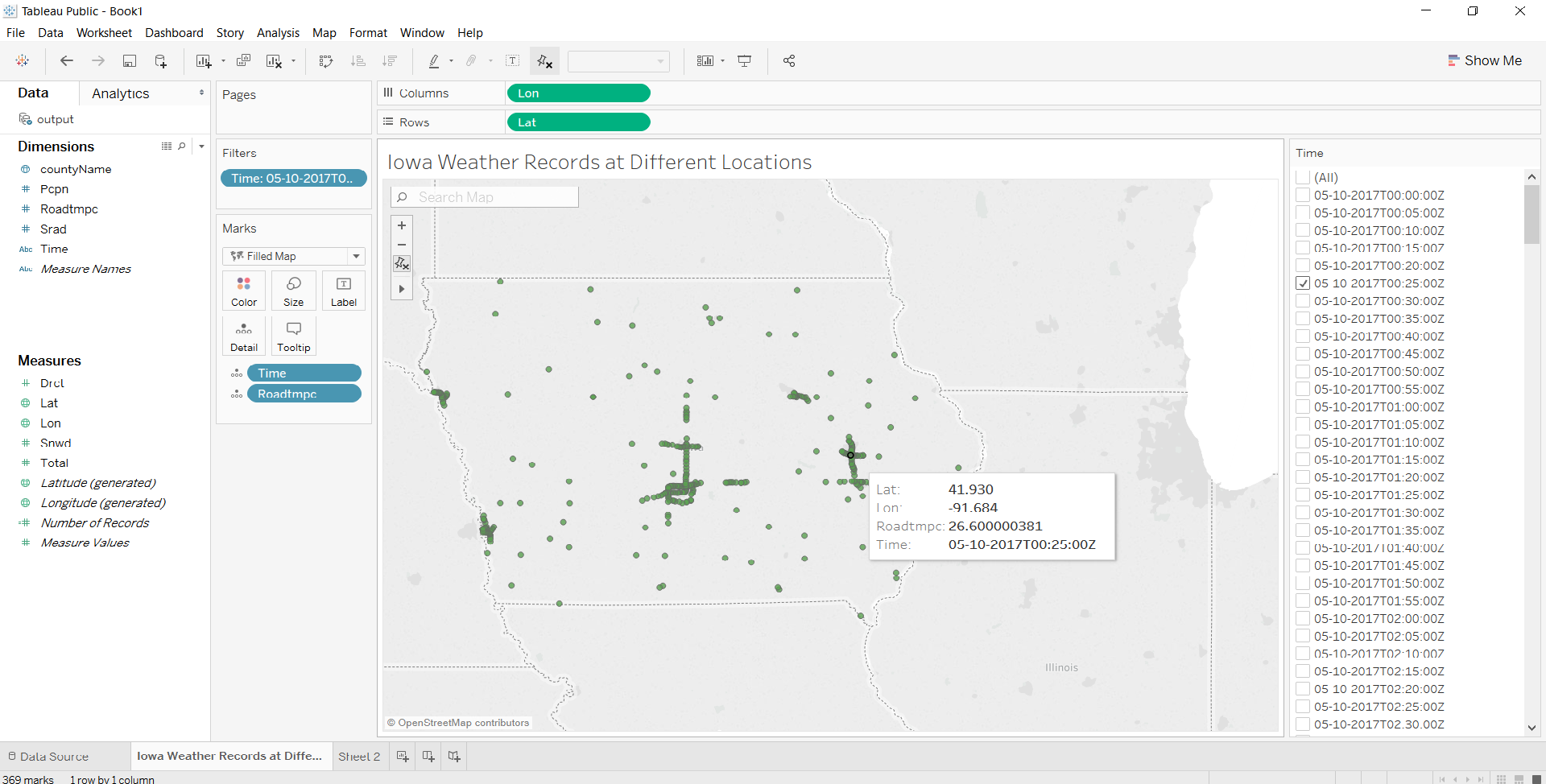}
	\caption{This Tableau dashboard shows the road temperatures in degree Celsius at different times of the day at different locations. We can select the time from the time selector panel on the right. And once hovering the marker we'll be able to see the road temperature at that location at that time. }   
	\label{fig:tableau}
\end{figure}

\begin{figure}[ht]
	\centering
	\includegraphics[width=0.8\textwidth]{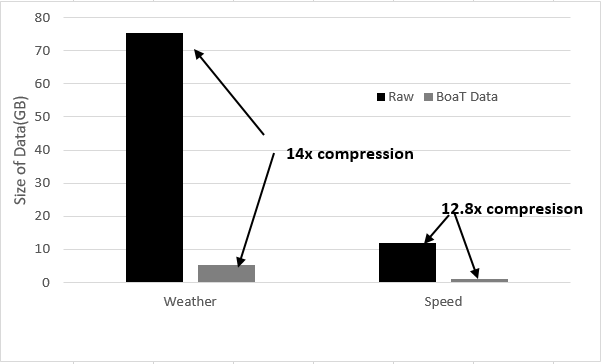}
	\caption{Reduction in data storage size in BoaT data infrastructure compared to the raw data}   
	\label{fig:storage}
\end{figure}

\subsection{Example Dashboard Visualization}
\label{subsec:dashboard}

BoaT query results can be used to create interactive visualizations and dashboards. 
To support this claim we present few examples of simple visualizations.

We present the query result from Task F.1 in a simple dashboard created using JavaScript and Google Map in  \figref{fig:highspeedlocations}. 
The markers show different speeding incident locations. 
Once a marker is clicked then the chart on the right side shows the number of vehicles recorded above 70 mph at that location. 
For example at location (41.97399057, -93.5702799) more than 150 vehicles 
were running at 71 mph on that day. 

We provide another visualization of task D.2 in \figref{fig:topten}. 
In this task, we find out the top ten counties where the average speed 
was higher than other counties on that day.
BoaT result can be easily imported to Tableau or other visualization softwares to show the results. 
To show an example of this we visualize the result of task A.5 in the tableau in 
\figref{fig:tableau}. 
DOTs and researchers who use visualization tools like tableau can directly 
benefit from the BoaT results.

\subsection{Storage Efficiency}
\label{subsec:storage}

For evaluating the benefit we compare raw data along with the data storage in BoaT. 
If we compress the raw data to reduce the size we would lose the performance of query therefore a compressed format is not desirable. 
But in BoaT, we can achieve the desired performance even after a huge reduction 
in the data size.
The language reads the objects according to the domain type and emits the result from the Hadoop nodes to produce the final result. 
For comparison, we used weather and speed data of one week for the state of Iowa. 
The weather data contains different weather information related grids at different locations at five minutes interval. 
The speed data contains the readings from Inrix sensors at 20 second intervals. 
The pre-processed raw weather data size 75.5 GB and the pre-processed raw speed data size is 12.07 GB. 
We took these datasets to generate an example BoaT dataset. 
On top of the raw weather and speed data, we add a lot more other data like county names of grids, county code, county names where the speed detector is located, road names of speed detectors. 
We collect some of this additional information from other metadata sources and some others using Google API. 
Even after adding a lot more additional data our generated BoaT dataset size is much smaller than the original raw data. 
The original 75.5 GB speed dataset is reduced to 5.38 GB in BoaT and the 
original 12.07 Gb speed dataset is reduced to 942 MB in BoaT as shown 
in \figref{fig:storage}

%% file: tables/querytable.tex
\newcounter{tasknum}
\newcommand{\task}{\stepcounter{tasknum}\thetasknum.\xspace}

\begin{sidewaysfigure}
\hspace*{-.250in}
	\rowcolors{2}{gray!25}{white}
	\begin{footnotesize}
		\begin{tabular}{|l|l||r|r|r||r|r|r|}
			\cline{3-8}
			\multicolumn{1}{c|}{} & \multicolumn{1}{c|}{} & \multicolumn{3}{c||}{\textbf{LOC}} & \multicolumn{3}{c|}{\textbf{RTime (sec)}}\\
			\hline
			\multicolumn{1}{|c|}{\textbf{Task}} & \multicolumn{1}{c|}{\textbf{Classification}} & \multicolumn{1}{c||}{\textbf{Java}} & \multicolumn{1}{c|}{\textbf{BoaT}} & \multicolumn{1}{c||}{\textbf{Diff}} & \multicolumn{1}{c|}{\textbf{Java}} & \multicolumn{1}{c|}{\textbf{BoaT}} & \multicolumn{1}{c|}{\textbf{Speedup}} \\
			\hline
			\multicolumn{8}{|c|}{\textbf{A. Temperature Statistics}} \\
			\hline
			\setcounter{tasknum}{0}%
			\task Compute the mean, standard deviation of temperature & Central Tendency &  84 & 10 & 8.4x & 465 & 22 & 21.14x \\
			\task Find the top ten counties with highest temperature & Rank & 90 & 17	& 5.29x	& 470 & 20 & 23.50x
			 \\
			\task Find the top ten counties with lowest temperature & Rank &  85 & 14	& 6.07x	& 489 &	25	& 19.56x
			 \\
			 \task Find the highest and lowest temperature in different Counties & Rank & 70 &	8	& 8.75x & 	485 & 	23	& 21.09x
			 \\
			 \task Find the correlation between Solar radiation and temperature & Correlation & 65 &	10 &	6.50x	& 460	& 18	& 25.56x
			 \\
			 \task Find the locations below a threshold temperature & Anomaly & 69	& 11 &	6.27x & 	498 & 	17 &	29.29x	 
			 
			 \\
			\hline
			\hline
			\multicolumn{8}{|c|}{\textbf{B. Wind Behavior }} \\
			\hline
			\setcounter{tasknum}{0}%
			\task Compute the mean, standard deviation of wind speed & Central Tendency &
			97 &	12	& 8.083x	& 2753	& 48	& 57.35x
			\\
			\task Find the top ten locations with higher wind speed & Rank
			 &  85 &	9	& 9.44x	& 2960	& 57	& 51.93x
			  \\
			\task Find the range across of wind different Counties & Rank
			 &  65 &	12 &	5.42x &	2793 &	45	& 62.07x
			  \\
			\task Find the correlation between temperature and wind speed & Correlation &
			  90 & 15 &	6.00x &	2743 &	43	& 63.79x \\
			\task Find the locations above a threshold weather speed & Anomaly
			 &  65 &	10	& 6.50x	& 2894	& 47	& 61.57x
			  \\

			\hline
			\hline
			\multicolumn{8}{|c|}{\textbf{C. Study of precipitation behavior}} \\
			\hline
			\setcounter{tasknum}{0}%
			\task Compute the mean, Standard deviation, Quantile of Precipitation & Central Tendency
			 &  65 &	17 &	3.82x &	2883	& 51	& 56.53x
			  \\
			\task Find the top ten Counties with high precipitation & Rank
			 &  35 &	10	& 3.50x &	2945	& 46	& 64.02x
			  \\
			\task Find the correlation of temperature with precipitation & Correlation & 80 &	9	& 8.89x	& 2980	& 45	& 66.22x
			  \\
			 \task Find the correlation of visibility with precipitation & Correlation &
			 80 & 	8	& 10.00x	& 2401	& 53	& 45.30x
			 \\
			 \task Find the locations where precipitation was above a threshold limit & Anomaly &
			 72 &	6 &	12.00x	& 2180	& 48	& 45.42x
			 \\

			\hline
			\hline
			\rowcolor{white}
			\multicolumn{8}{|c|}{\textbf{D. Study of Speed}} \\
			\hline
			\rowcolor{gray!25}
			\setcounter{tasknum}{0}%
			\task Compute the mean, standard deviation of speed at different locations & Central Tendency &
			 100	& 17	& 5.88x	& 860	& 31	& 27.74x
			  \\
			
			\rowcolor{gray!25}
			\task Find the top ten counties with higher average speed & Rank
			 &  80 &	9	& 8.89x	 & 815	& 25	& 32.60x
			  \\
			\rowcolor{white}
			\task Find the road names with higher average Speed? & Rank
			 &  75 &	11	& 6.82x	& 810	& 27	& 30.00x
			  \\
			\rowcolor{gray!25}
			\task Find the county with maximum and minimum average speed & Rank
			 &  90 &	8 &	11.25x	& 986	& 35	& 28.17x
			  \\
			  
			\hline
			\hline
			\rowcolor{white}
			\multicolumn{8}{|c|}{\textbf{E. Effect of weather on speed}} \\
			\hline
			\rowcolor{gray!25}
			\setcounter{tasknum}{0}%
			\task Find the Correlation between speed and precipitation & Correlation 
			 &  150 &	13	& 11.54x &	4230 &	130 &	32.54x			 
			  \\
	
			\rowcolor{white}
			\task Find the Correlation between Speed and Visibility & Correlation
			 &  150	& 13	& 11.54x	& 4213	& 132	& 31.92x
			  \\
			  \rowcolor{gray!25}
			  \task Find which weather parameter is more correlated with speed & Correlation
			  &  165 &	17 &	9.71x &	4560	& 135	& 33.78x 			  
			  \\
			  \hline
			  \hline
			  \rowcolor{white}
			  \multicolumn{8}{|c|}{\textbf{F. Speeding Violations}} \\
			  \hline
			  \rowcolor{gray!25}
			  \setcounter{tasknum}{0}%
			  \task Find the number of vehicles running above a speed limit in different locations
			   & Anomaly 
			  &  80	& 12	& 6.67x	& 980	& 28	& 35.00x			  		 
			  \\

			  \rowcolor{white}
			  \task What is the percentage of vehicles above reference speed at different locations?
			   & Anomaly
			  &  95	& 15	& 6.33x & 	953 & 	23	& 41.43x			  
			  \\
			  
			  \rowcolor{gray!25}
			  \task Find the number of under speeding vehicles at different locations
			  & Anomaly
			  &  80 &	12 &	6.67x	& 910	& 27	& 33.70x			  
			  \\
			\hline
		\end{tabular}
	\end{footnotesize}
	\caption{Example of BoaT programs to compute different tasks on transportation data}
	\label{tbl:app-result}
\end{sidewaysfigure}
\newcommand{\BoaMean}{79\xspace}
\newcommand{\BoaMax}{753\xspace}
\newcommand{\BoaMin}{24\xspace}
\newcommand{\JavaMean}{9607\xspace}
\newcommand{\JavaMax}{86400\xspace}
\newcommand{\JavaMin}{609\xspace}
\newcommand{\BoaLOCMean}{6\xspace}
\newcommand{\BoaLOCMax}{30\xspace}
\newcommand{\BoaLOCMin}{2\xspace}
\newcommand{\JavaLOCMean}{70\xspace}
\newcommand{\JavaLOCMax}{180\xspace}
\newcommand{\JavaLOCMin}{32\xspace}
\newcommand{\BoaImproveMean}{122\xspace}
\newcommand{\BoaImproveMin}{24\xspace}
\newcommand{\BoaImproveMax}{569\xspace}
\newcommand{\BoaLOCImproveMean}{12\xspace}
\newcommand{\BoaLOCImproveMin}{6\xspace}
\newcommand{\BoaLOCImproveMax}{22\xspace}

%% file: examples/example1.tex
\begin{figure}[ht]
\centering
\begin{lstlisting}
	p: County = input;
	max: output maximum(1)[string] of string weight float;
	min: output minimum(1) of string weight float;
	foreach(i : int; def(p.grid[i])){
		weatherRoot := getweather(p.grid[i],"5-11-2017");
		foreach(j : int; def(weatherRoot.weather[j])){
			max << p.countyName weight weatherRoot.weather[j].tmpc;
			min << p.countyName weight weatherRoot.weather[j].tmpc;		   	   
		}
	}
\end{lstlisting}
\caption{Task A.4: Find the highest and lowest temperature in different counties}
\label{fig:example1}
\end{figure}

%% file: examples/example2.tex
\begin{figure}[ht]
\centering
\begin{lstlisting}
	p: County = input;
	average : output mean[string] of int;
	stdev : output stdev[string] of int;
	visit(p, visitor {
		before n: Grid -> {
			speedRoot := getspeed(n,"5-11-2017");
			foreach(s : int; def(speedRoot.speeds[s])) {
				average[p.countyName] << speedRoot.speeds[s].speed;
				stdev[p.countyName] << speedRoot.speeds[s].speed;
			}
		}
	});
\end{lstlisting}
\caption{Task D.1: Compute the mean and standard deviation of speed at different locations}
\label{fig:example2}
\end{figure}

%% file: conclusion.tex
\section{Conclusion and Future Work}
\label{sec:conclusion}
%

Big Data-driven transportation engineering is ripe with potential to make a significant impact. However, it is hard to get started today. 
In this work, we have proposed BoaT, a transportation-specific Big Data programming language that is designed from the ground up to 
simplify expressing data analysis task by abstracting away the tricky
details of data storage strategies, parallelization, data aggregation, etc.
We showed the utility of our new approach, as well as its scalability advantages. Our future work will try out more application as well as create a web-based infrastructure so that others can also take advantage
of BoaT's facilities. 